\documentclass[onecolumn,10pt]{revtex4}
\usepackage{amsmath,amsthm,amssymb}

\renewcommand{\>}{\rangle}
\newcommand{\C}{\mathcal{C}}

\newcommand{\e}{\epsilon}
\newcommand{\Wm}{W^{\{m\}}}
\newcommand{\Gm}{G^{\{m\}}}
\newcommand{\Qm}{Q^{\{m\}}}
\newcommand{\E}{E_{\epsilon}}
\newcommand{\F}{\mathbb{F}}

\theoremstyle{plain} 
\newtheorem*{theorem*}{Theorem}
\newtheorem*{conclusion*}{Conclusion}

\begin{document}

\title{Probabilistic quantum error correction}
\author{Jesse Fern}\email{jesse@math.berkeley.edu}
 \affiliation{Department of
   Mathematics, University of California, Berkeley, CA
   94720-3840.}
\author{John Terilla}\email{jterilla@math.sunysb.edu} \affiliation{Department of
   Mathematics, State University of New York, Stony Brook, NY
   11794-3651.}

\begin{abstract}There are well known necessary and sufficient
  conditions for a quantum code to correct errors.  We
  study weaker conditions under which a quantum code
  may correct errors with probabilities that may be less than one.  We
  work with stabilizer codes and as an application study how the nine
  qubit code, the seven qubit code, and the five qubit code perform
  when there are errors on more than one qubit.  As a second application, we discuss the concept of syndrome quality and use it to 
suggest a way that quantum error correction can be practically improved.
\end{abstract}

\maketitle
\tableofcontents
\acknowledgements{This research was conducted 
at the State University of New York, Stony Brook and was funded by a VIGRE grant.}

\section{Introduction}
A quantum $[[n,k,d]]$ code can correct, with
probability $1$, arbitrary errors on $t$ qubits, provided $2t<d$.  However,
it is possible that an $[[n,k,d]]$ code will have a nonzero probability
of correcting arbitrary errors affecting $t$ qubits even if $2t\geq d.$
This possibility figures more prominently in quantum codes than classical ones
because of the probabilistic nature of quantum measurement.
This paper examines, within the context of stabilizer codes, the
conditions under which a code may correct errors on more qubits than
it is guaranteed to fix and gives a framework in which to compute the
probabilities that an arbitrary error will be corrected.  
Similar ideas are developed in the two articles \cite{ABKL1,ABKL2} on 
error detection.
As an application, we supply an analysis of these probabilities for the most
famous one qubit codes, comparing well known
$[[9,1,3]]$, $[[7,1,3]]$, and
$[[5,1,3]]$ codes.  We examine these codes further, assuming the
independent error model with the probability of $p$ for an error occurring on any one qubit.  Interestingly, for some $p$, the $[[9,1,3]]$ code is more effective than the $[[7,1,3]]$ code, and for other $p$ it is not (see the plot in section (\ref{figuresection}).)

During the course of quantum error correction, an error syndrome is measured and the correction procedure continues dependent on this measurement.  As a second application of probabilistic error correction, we analyze the likelyhood that error correction will succeed given that a particular syndrome is measured.  This likelyhood, which we call the syndrome quality, may enhance the effectiveness of a quantum information process for which quantum error correction plays a role.  For example, it is conceivable that a quantum information process may benefit from aborting
a subroutine if at some point a syndrome of especially low quality is measured.

\section{Stabilizer codes}
In this section, we establish our notation.  For an introduction
to stabilizer codes, see \cite{G2,G1}, or section 10.5 in \cite{CN}.    
The Pauli group on $n$ qubits
is a subgroup of linear operators on
$W:=\left(\mathbb{C}^2\right)^{\otimes n}$ defined by
\begin{equation}
G^{n} = \{ \pm I, \pm X, \pm Y, \pm Z, \pm iI, \pm iX, \pm iY, \pm iZ\} ^{\otimes n}
\end{equation}
where
\begin{equation}
X= \begin{pmatrix}0 & 1 \\ 1 & 0 \end{pmatrix}, \quad
Y=\begin{pmatrix}0 & 1 \\ -1 & 0 \end{pmatrix}, \quad Z=
\begin{pmatrix}1 & 0 \\ 0 & -1 \end{pmatrix}.
\end{equation}
Any linear operator on $W$ can be written as a complex 
linear combination of
elements of $G^n$.  Note that any two elements of $G^n$ either commute
or anticommute.

Suppose $S$ is an abelian subgroup of $G^n$ with $-I \notin S$, and
that
$g_1,g_2,\ldots g_r$ are independent generators of $S$.  It follows that every element $g\in S$ satisfies $g^2=I$ and has eigenvalues $\pm 1.$
The subset $C_S\subseteq W$ defined by
\begin{equation} 
C_S = \{{|\psi\> \in W} \text{ so that } g |\psi\> = |\psi\> \text{ for all
}g\in S\}
\end{equation}
is a vector space of dimension $2^{k}$, where $k=n-r$.  The space $C_S$
can be regarded as the image of a space of $k$ qubits, embedded as a code space of $W$.  The space $C_S$ is called a stabilizer code, 
and the group $S$ is called the stabilizer of $C_S$.

\subsection{Error detection}
For $|\psi\> \in C_S$, one imagines that the state $|\psi\>$ is
transformed to $E|\psi\>$ by some linear operator $E:W \to
W$ representing an error.  Error detection begins by measuring the
observables $g_1,g_2,\ldots, g_r$.  Once the measurements have been made, 
one has an \emph{error syndrome} $\e:\hom(W,W)\rightarrow \F_2^r$, defined
by
\begin{equation}
\e(E)= (\e_1(E),\e_2(E),\ldots,\e_r(E)),
\end{equation}
where $\e_i(E)=0$ if the measurement of $g_i$ projects the state
$E|\psi\>$ onto the $+1$ eigenspace of $g_i$, and $\e_i(E)=1$ if the
measurement of $g_i$ projects the state $E|\psi\>$ onto the $-1$
eigenspace of $g_i$.
\paragraph{Error syndrome for $G^n$.}
The error syndrome is obtained after a sequence of measurement of
$E|\psi\>$ and hence is, in general, determined by the error $E$ only
probabilistically.  However, if the error $E=g\in G^n$, then 
$g|\psi\>$ is an
eigenvector of each $g_i$, for any $|\psi\>\in C_S$, so the measurement outcomes are determined
and do not disturb $g|\psi\>$.  The error syndrome $\e(g)$, in this
case, will always be given by
\begin{equation} 
\e_i(g) =
 \begin{cases}
   0 & \text{ if }gg_i=g_ig, \\
   1 & \text{ if }gg_i=-g_ig.
 \end{cases}
\end{equation}
Note that for any $g,h \in G^n$,
\begin{equation}
\e(gh)=\e(g)+\e(h).
\end{equation}
In particular, if $h\in \C(S)$, the centralizer of $S$, then
$\e(gh)=\e(g)$ since $\e(h)=(0,0,\ldots,0)$.  Therefore, the error
syndrome $\e$ is constant on the equivalence classes in $G^n$ defined
modulo the normal subgroup $\C(S).$ Furthermore, different equivalence classes are given
by different syndromes.  We form the quotient group
\begin{equation}
Q_S=G^n / \C(S).
\end{equation}
Each element of $Q_S$ is determined by a syndrome $\e \in
\F_2^r$ and consists of the set 
\begin{equation}
G_{\e}=\{g \in G^n: \e(g)=\e\}.
\end{equation}

\paragraph{Decomposition of an arbitary error under measurement.}
Let us fix a subset $\{m\}=\{m_1,m_2,\ldots, m_l\} \subseteq
\{0,1,\ldots ,n\}$.  Let $\Wm$ be the space spanned by the qubits in
the $m_1, m_2, \ldots, m_l$ factors of $W$.  Consider a linear
operator $E:\Wm\to \Wm$; i.e., an error acting only the qubits $m_1,
m_2, \ldots, m_l$.  We define a subgroup $\Gm$ of $G^n$ by
\begin{equation}
\Gm=\hom\left(\Wm,\Wm\right)\cap G^n.
\end{equation}
Then an arbitrary $E$ can be expressed as a linear combination of the
elements of $\Gm$:
\begin{equation} 
E = \sum_{g \in \Gm} a_g g, \text{ for some }a_g \in \mathbb{C}.
\label{err}
\end{equation}
If the system is in the state $E|\psi\>$ for some $|\psi\>\in C_S$,
then measuring $g_1,g_2, \ldots,g_r$ yields a syndrome $\e$ and
projects $E|\psi\>$ onto the state $\E|\psi\>$ consisting of only
those terms $\Gm$ which give the same syndrome:
\begin{equation}
\E=\sum_{g \in \Gm_{\e}}a_g g,
\end{equation}
where $\Gm_\e$ is the subgroup of $\Gm$ defined by 
\begin{equation}
\Gm_{\e}:=\Gm \cap G_{\e}.
\end{equation}
It will be convenient to denote by $\Qm_S$ the quotient
group
\begin{equation}
\Qm_S=\Gm / \left( \Gm \cap \C(S)\right)
\end{equation}
whose elements are the equivalence classes $\Gm_{\e}.$

\subsection{Error correction}
\paragraph{Error correcting function.}
The outcome of the error detection is the syndrome.  Recovery from an
error is accomplished by using an \emph{error correction
  function} $\phi : \F_2^r \to G^n$, which assigns an operator
$\phi(\e)$ to each syndrome $\e \in \F_2^r$.  The two step
detection/recovery procedure amounts to
\begin{equation}
E|\psi\> \mapsto \phi(\e)\E|\psi\>
\end{equation}
and is considered successful if $\phi(\e)\E|\psi\>=|\psi\>$.  
We assume that an error correction function has been given.  Traditionally, the error correction function is not considered part of the data of the code.  The reason, in part at least, is that the usual approach to quantum error correcting codes treats errors that are not correctible with one-hundred percent probability as uncorrectible.   
The following restatement of theorem 10.8 from \cite{CN} summarizes the usual approach:
\begin{theorem*}
Given a set of errors $\{E_j\}_{j=1}^s\subset G^n$, there exists an error
correcting function $\phi$ so that, for all errors of the form
\begin{equation}
E=\sum_{j=1}^s a_j E_j \text{ for any }a_j \in \mathbb{C},
\end{equation} 
the error detection/correction procedure succeeds with probability one, if and only if $E_j^\dagger E_k \notin C(S)\setminus S$ for all $j,k$.
\end{theorem*}  
Any such error correcting function will have the same values on the syndromes $\{\e(E_j)\}$ and serve equally well in correcting the errors $E$ that are correctable with probability one.  However, in general, there is ambiguity in choosing the values of $\phi$ on $\F_2^n \setminus \{\e(E_j)\}$, and unless one wishes to throw out all 
instances where the syndrome does not lie within $\{\e_{E_j}\}$, one must make some choices.

For a nondegenerate code, every nonzero
syndrome corresponds exactly to an error that can be corrected with probability one.  In the nondegenerate $[[5,1,3]]$ code, every syndrome corresponds to either an $X,Y,$ or $Z$ error on one of the
five qubits, and the error correction function is completely determined by
this correspondence.  Certainly, effectiveness of
a degenerate code, such as Shor's $[[9,1,3]]$ code, can be
improved with a carefully chosen error correction function. 

In addition, one imagines that the choice of error function should depend on a particular model for errors and should be chosen to maximize the
probability of recovery, regardless of whether the code is
degenerate or not. The independent error model is a reasonable model to consider for a quantum computer.  In this model, one assumes that the qubits interact only with their local environment.  Thus, one has a parameter $p\in [0,1]$ representing the probability that an error will occur on any particular qubit and one assumes that errors are uncorrelated.  

\subsection{Probabilistic error correction}\label{mainresults}

For an $[[n,k,d]]$ code, it is important to understand how likely it will correct errors on $t$ qubits with $2t\geq d.$   
For example, both the $[[5,1,3]]$ code and the $[[7,1,3]]$ codes will correct an arbitrary one qubit error with probability $1$.  As we show in section \ref{examplesection}, the $[[5,1,3]]$ code will correct an arbitrary two-qubit error with probability $\frac{7}{16}$ and the $[[7,1,3]]$ code will correct an arbitrary two qubit error with probability $\frac{9}{16}$.

Consider a general code with stabilizer $S=\<g_1,\ldots, g_r\>\subset G^n$ and an arbitrary error $E:\Wm \to \Wm$ affecting a state 
$|\psi\>\in C_S$ at a given set $\{m\}=\{m_1,\ldots,m_l\}$ of qubits.  Write $E=\sum_{g \in \Gm}a_g g$.  A syndrome measurement
$\e\in \F_2^r$ projects $E|\psi\>$ to $E_\e |\psi\>$ with $E_\e=\sum_{g \in \Gm_\e}a_g g$.  Then, the error correction will succeed provided
\begin{equation}
\phi(\e)\E=\mathrm{Id}_{C_S} \Leftrightarrow \phi(\e)\E\in S.
\end{equation}
For randomly occurring $E$, the coefficients $a_g \neq 0$, so $\phi(\e)\E\in S$ if and only if $\phi(\e)g \in S$ for each $g \in \Gm_\e$.
Thus, we arrive at the following
\begin{conclusion*}
An randomly occuring $E:\Wm \to
\Wm$ will be corrected successfully if and only if 
a measurement of $g_1,\ldots, g_r$ yields a syndrome $\e$ satisfying \begin{equation}\label{condition1}
\phi(\e)g \in S \text{ for every }g \in \Gm_\e.
\end{equation}
\end{conclusion*}

One should consider the condition in this conclusion as a condition
for both the set $\Gm_\e$, depending on the vector $\e\in \F_2^n$ and the set $\{m\}\subseteq \{1,\ldots,n\}$.  We give two applications:

\begin{enumerate}
\item We apply our conclusion to analyze the probabilities that an error will be corrected given that it affects $t$ qubits in specified locations.  This amounts to holding $\{m\}$ fixed and counting the $\e$ that satisfy condition (\ref{condition1}).  Then, we give the probability that an error affecting $t$ qubits in unspecified locations will be corrected.  This analysis is important in order to understand how effective a stabilizer code really is, assuming an independent model of quantum errors.  
\item Then, in section \ref{syndromesection}, we apply the conclusion to  introduce \emph{syndrome quality}.
We study the probabilities that an error will be corrected given that a particular syndrome is measured, but an unknown number of qubits is specified.  This amounts to holding $\e$ fixed and counting the $\{m\}$ that satisfy condition (\ref{condition1}).  This syndrome quality is important as it can be used to improve the implementation of a quantum code via classical subroutines that abort \emph{after} syndrome mesaurement when successful error recovery is particularly unlikely.
\end{enumerate}

In order to determine the probability that a randomly occuring error $E$ affecting a given set $\{m\}=\{m_1,\ldots,m_l\}$ of qubits will be corrected, we regard condition (\ref{condition1}) as a condition on the elements of $\Qm_\e$ and count to find:
\begin{equation} 
\mathrm{Prob}(E \text{ will be corrected})=
\frac{\left|\left \{\Gm_\e \in \Qm_S  | \phi(\e)g \in S \text{ for every }g \in \Gm_\e \right\}\right|}{\left|\Qm_S\right|}.
\end{equation}
To study errors that affect $t$ qubits in unspecified
locations we introduce a function $f_S$.  
For each integer $t=0,1,\ldots,n$, we define $f_S(t)$ to be the probability that a random error affecting any $t$ of the $n$ qubits will be corrected.
One has:
\begin{equation}\label{f-function}
f_S(t)=\frac{1}{\binom{n}{t}} \sum_{\{m\}}\frac{ \left|\left \{\Gm_\e \in \Qm_S  | \phi(\e)g \in S \text{ for every }g \in \Gm_\e \right\}\right| }{\left|\Qm_S\right|}.
\end{equation}
where the sum is over all $t$ element subsets $\{m\}=\{m_1,\ldots,
m_t\}$ of $\{1,\ldots, n\}$.

\section{Probabilistic analysis of three 1-qubit codes}\label{examplesection}

Throughout this section, we compactify notation by omitting tensor signs.  For example, we abbreviate $X \otimes Z \otimes Z \otimes X \otimes I$ by $XZZXI$.

\subsection{A $[[5,1,3]]$ code}
In 1996, two groups \cite{BDSW,LMPZ} independently discovered the same $[[5,1,3]]$
code.   Let us denote the stabilizer of this code by $S_5=\<g_1,g_2,g_3,g_4\>\subset G^5$.  We specify generators:
\begin{gather}
g_1=XZZXZ\\
g_2=IXZZX\\
g_3=XIXZZ\\
g_4=ZXIXZ.
\end{gather}
The error correcting function $\phi:\F_2^4 \to G^5$ is uniquely defined (up to multiples of stabilizers) once it has been chosen to correct all $0$ and $1$ qubit errors:

\begin{equation}
\begin{aligned}
\phi(0000)&=IIIII & \phi(0001)&=XIIII\\ 
\phi(0010)&=IIZII & \phi(0011)&=IIIIX\\ 
\phi(0100)&=IIIIZ & \phi(0101)&=IZIII\\ 
\phi(0110)&=IIIXI & \phi(0111)&=IIIIY\\ 
\phi(1000)&=IXIII & \phi(1001)&=IIIZI\\ 
\phi(1010)&=ZIIII & \phi(1011)&=YIIII\\ 
\phi(1100)&=IIXII & \phi(1101)&=IYIII\\ 
\phi(1110)&=IIYII & \phi(1111)&=IIIYI. 
\end{aligned}
\end{equation}

It is straightforward to check that this code corrects all $0$ and $1$ qubit errors with probability $1$, so the function $f_{S_5}$ (see equation (\ref{f-function})) satisfies:
\begin{equation}
f_{S_5}(0) = f_{S_5}(1) = 1.
\end{equation}
On any given $2$ qubit space $W^{\{j,k\}}$, $IIIII$ is the only element that is in the $G^{\{j,k\}}\cap \C(S)$ so we correct $E:W^{\{j,k\}} \to W^{\{j,k\}}$ if and only $E_\epsilon=\phi(\e)$, where $\epsilon$ is the error syndrome measured; that is, if after measurement $E_\e$ affects at most one qubit.   Given $E$, the measurement $\e(E)$ projects $E$ onto $E_\e$ which acts on each of the affected qubits as $\{I,X,Y,Z\}$ with equal probability. This yields,
\begin{equation}
f_{S_5}(2)=7/16.
\end{equation}
For any $\{m\}$ with $|\{m\}|>2$, condition (\ref{condition1}) always fails, so no error on more than two qubits will be corrected.

\paragraph{Example.}Here is a simple example.  
Consider the error $E=\cos(\theta)IZIII-i\sin(\theta)YZIII$.  (This example is quite artificial.  A randomly occurring error affecting qubits $1$ and $2$ would be a sum of sixteen nonzero terms, not just two.) 
When the syndrome is measured,
one obtains
\begin{gather}
\e(E)=(0101) \text{ with probability }\cos^2(\theta) \\
\e(E)=(1110) \text{ with probability }\sin^2(\theta).
\end{gather}
If $\e(E)=(0101)$, then $E$ is projected to $E_{(0101)}=IZIII$ and $\phi(0101)=IZIII$ is applied to $E_{(0101)}|\psi\>=IZIII|\psi\>$, yielding $|\psi\>$ and correcting the error.  If $\e(E)=(1110)$, then $E$ is projected to $E_{(1110)}=YZIII$ and $\phi(1110)=IIYII$ is applied to $E_{(1110)}|\psi\>=YZIII|\psi\>$, erroneously yielding the state $YZYII|\psi\>$.

\subsection{A $[[7,1,3]]$ code}
By adapting a classical Hamming code, Steane produced a $[[7,1,3]]$
quantum code \cite{St1}.  It can be viewed as a stabilizer code with 
stabilizer group $S_7=\<g_1,g_2,g_3,g_4,g_5,g_6\>$.  We specify generators:
\begin{equation}
\begin{aligned}
g_1&=IIIXXXX\\
g_2&=IXXIIXX\\
g_3&=XIXIXIX\\
g_4&=IIIZZZZ\\
g_5&=IZZIIZZ\\
g_6&=ZIZIZIZ.
\end{aligned}
\end{equation}
This code has its origin as a self dual Hamming code, so it corrects $X$ and $Z$ errors separately.  For this reason, 
it is possible to define the error correcting function $\phi$ so that it corrects $X$ qubit errors and $Z$ qubit errors on distinct qubits simultaneously.  This defines $\phi$ completely and every error syndrome corresponds to an error affecting at most two qubits.  It can be checked that for an independent error model, the error correcting function thus defined is optimal.

The probabilities $f_{S_7}(t)$ are: 
\begin{equation}\label{f_{S_7}}
f_{S_7}(0)=1, \quad f_{S_7}(1)=1, \quad f_{S_7}(2)=\frac{9}{16}, \quad f_{S_7}(3)=\frac{5}{16}, \quad f_{S_7}(4)=\frac{5}{64}, \quad f_{S_7}(5)=f_{S_7}(6)=f_{S_7}(7)=0.
\end{equation}
It is straightforward, but tedious, to compute these probabilities.  
Let us illustrate a typical computation.  It is interesting that $f_{S_7}(4)\neq 0$, so we consider an example showing how it may happen 
that Steane's one qubit code could correct errors affecting four of the seven qubits.

\paragraph{Example}
Suppose that an error affecting qubits $1,3,5,$ and $7$ occurs and that a syndrome of $(000111)$ is measured.  We find that 
\begin{equation}
G^{\{1,3,5,7\}}_{(000111)} = \{ XIXIXII, IIIIIIX, YIYIYIZ, ZIYIZIY \}.
\end{equation}
These are all the elements of the Pauli group that act as the identity on qubits $2,4,$ and $6$ that commute with $g_1,g_2,g_3$ and anticommute with $g_4,g_5,g_6.$  After measuring the observables $g_1,\ldots, g_6$, an arbitrary error affecting qubits $1,3,5,$ and $7$, is projected to a linear combination of the four elements of $G^{\{1,3,5,7\}}_{(000111)}.$
One checks that $\phi(000111)=IIIIIIX$ (a bit flip on the seventh qubit produces a syndrome of $(000111)$, and $\phi$ is selected so that such an error is corrected).  We check
\begin{align}
\phi(000111)\circ XIXIXII &= IIIIIIX \circ XIXIXII = XIXIXIX =g_3 \in S \\
\phi(000111)\circ IIIIIIX &= IIIIIIX \circ IIIIIIX = IIIIIII \in S \\
\phi(000111)\circ YIYIYIZ &= IIIIIIX \circ YIYIYIZ = YIYIYIY =g_6g_3 \in S \\
\phi(000111)\circ ZIZIZIY &= IIIIIIX \circ ZIZIZIY = ZIZIZIZ=g_6 \in S.
\end{align}
Thus, condition (\ref{condition1}) is satisfied for $G^{\{1,3,5,7\}}_{(000111)}$, implying that $f_{S_7}(4)\neq 0$.

As indicated in equation (\ref{f_{S_7}}),
$f_{S_7}(4)=\frac{5}{64}$.  So, it is more likely for $|\{m\}|=4$ 
that $\Gm_\e$ will not satisfy the condition (\ref{condition1}).  
For instance, one can check that $IIXXIII \in G^{\{1,2,3,4\}}_{(000111)}$ and $\phi(000111)\circ IIXXIII = IIXXIIX =\notin S$.  Also, one has $IIIIXIX \in G^{\{1,3,5,7\}}_{(000010)}$ and $\phi(000010)\circ IIIIXIX = IXIIIII \circ IIIIXIX = IXIIXIX \notin S$.  So neither $G^{\{1,2,3,4\}}_{(000111)}$ nor $G^{\{1,3,5,7\}}_{(000010)}$ satisfy (\ref{condition1}).

\subsection{A $[[9,1,3]]$ code}
The first quantum error correcting code was constructed in 1995 by
Shor \cite{Shor1}.  It is a $[[9.1.3]]$ stabilizer code (other $[[9,1,3]]$ codes have been discovered since then)
and can be described as the code stabilized by the group $S_9=\{g_1,\ldots, g_8\}\subset G^9$ with
\begin{equation}
\begin{aligned}
g_1&=ZZIIIIIIII\\
g_2&=IZZIIIIII\\
g_3&=IIIZZIIII\\
g_4&=IIIIZZIII\\
g_5&=IIIIIZZI\\
g_6&=IIIIIIIZZ\\
g_7&=XXXXXXIII\\
g_8&=XIIIXXXXX.
\end{aligned}
\end{equation}

This code was constructed to correct any error on a
single qubit and one should select the error correcting function
$\phi$ in line with this intention.  However, this does not completely determine $\phi$.  We used a computer program to search the error correction functions to find the optimal $\phi$, assuming the independent model for errors.  We found that

\begin{equation}
f_{S_9}(0)=h_{S_9}(1)=1, \quad f_{S_9}(2)=\frac{51}{64}, \quad f_{S_9}(3)=\frac{45}{112}, \quad f_{S_9}(4)=\frac{15}{128}, \quad f_{S_9}(t)=0, \quad t>4.
\end{equation}

\subsection{Comparison plot}\label{figuresection}

Suppose that the probability of an error occuring on a given
qubit is $p$ and that errors occur independently.  
Then the probability that a code with stabilizer $S$ 
will succeed can be assembled as a
function of $p$:
\begin{equation}\label{h-function}
h_S(p)= \sum_{t=0}^n f_S(t)\binom{n}{t}p^t(1-p)^{n-t}.
\end{equation}
The plots of $h_S(p)$, for $S=S_5, S_7,$ and $S_9$ are as follows:

\bigskip

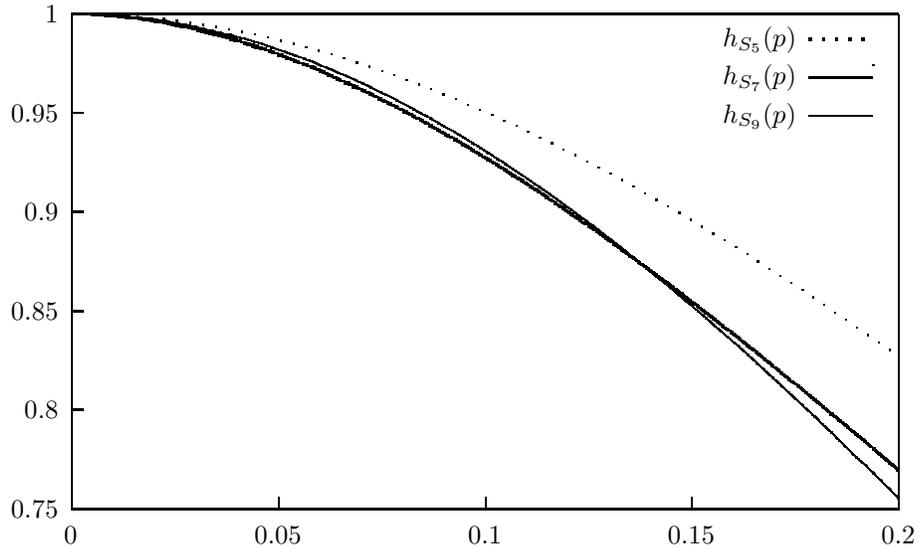
\begin{center}
\setlength{\unitlength}{0.240900pt}
\ifx\plotpoint\undefined\newsavebox{\plotpoint}\fi
\sbox{\plotpoint}{\rule[-0.200pt]{0.400pt}{0.400pt}}%
\begin{picture}(1500,900)(0,0)
\font\gnuplot=cmr10 at 10pt
\gnuplot
\sbox{\plotpoint}{\rule[-0.200pt]{0.400pt}{0.400pt}}%
\put(120,82){\makebox(0,0)[r]{$0.75$}}
\put(140.0,238.0){\rule[-0.200pt]{4.818pt}{0.400pt}}
\put(120,238){\makebox(0,0)[r]{$0.8$}}
\put(140.0,393.0){\rule[-0.200pt]{4.818pt}{0.400pt}}
\put(120,393){\makebox(0,0)[r]{$0.85$}}
\put(140.0,549.0){\rule[-0.200pt]{4.818pt}{0.400pt}}
\put(120,549){\makebox(0,0)[r]{$0.9$}}
\put(140.0,704.0){\rule[-0.200pt]{4.818pt}{0.400pt}}
\put(120,704){\makebox(0,0)[r]{$0.95$}}
\put(120,860){\makebox(0,0)[r]{$1$}}
\put(140,41){\makebox(0,0){0}}
\put(465.0,82.0){\rule[-0.200pt]{0.400pt}{4.818pt}}
\put(465,41){\makebox(0,0){$0.05$}}
\put(790.0,82.0){\rule[-0.200pt]{0.400pt}{4.818pt}}
\put(790,41){\makebox(0,0){$0.1$}}
\put(1114.0,82.0){\rule[-0.200pt]{0.400pt}{4.818pt}}
\put(1114,41){\makebox(0,0){$0.15$}}
\put(1439,41){\makebox(0,0){$0.2$}}
\put(140.0,82.0){\rule[-0.200pt]{312.929pt}{0.400pt}}
\put(1439.0,82.0){\rule[-0.200pt]{0.400pt}{187.420pt}}
\put(140.0,860.0){\rule[-0.200pt]{312.929pt}{0.400pt}}
\put(140.0,82.0){\rule[-0.200pt]{0.400pt}{187.420pt}}

\put(140,860){\usebox{\plotpoint}}
\put(166,858.67){\rule{3.132pt}{0.400pt}}
\multiput(166.00,859.17)(6.500,-1.000){2}{\rule{1.566pt}{0.400pt}}
\put(140.0,860.0){\rule[-0.200pt]{6.263pt}{0.400pt}}
\put(192,857.67){\rule{3.373pt}{0.400pt}}
\multiput(192.00,858.17)(7.000,-1.000){2}{\rule{1.686pt}{0.400pt}}
\put(206,856.67){\rule{3.132pt}{0.400pt}}
\multiput(206.00,857.17)(6.500,-1.000){2}{\rule{1.566pt}{0.400pt}}
\put(219,855.17){\rule{2.700pt}{0.400pt}}
\multiput(219.00,856.17)(7.396,-2.000){2}{\rule{1.350pt}{0.400pt}}
\put(232,853.67){\rule{3.132pt}{0.400pt}}
\multiput(232.00,854.17)(6.500,-1.000){2}{\rule{1.566pt}{0.400pt}}
\put(245,852.67){\rule{3.132pt}{0.400pt}}
\multiput(245.00,853.17)(6.500,-1.000){2}{\rule{1.566pt}{0.400pt}}
\put(258,851.17){\rule{2.700pt}{0.400pt}}
\multiput(258.00,852.17)(7.396,-2.000){2}{\rule{1.350pt}{0.400pt}}
\put(271,849.17){\rule{2.700pt}{0.400pt}}
\multiput(271.00,850.17)(7.396,-2.000){2}{\rule{1.350pt}{0.400pt}}
\put(284,847.17){\rule{2.700pt}{0.400pt}}
\multiput(284.00,848.17)(7.396,-2.000){2}{\rule{1.350pt}{0.400pt}}
\multiput(297.00,845.95)(2.918,-0.447){3}{\rule{1.967pt}{0.108pt}}
\multiput(297.00,846.17)(9.918,-3.000){2}{\rule{0.983pt}{0.400pt}}
\put(311,842.17){\rule{2.700pt}{0.400pt}}
\multiput(311.00,843.17)(7.396,-2.000){2}{\rule{1.350pt}{0.400pt}}
\multiput(324.00,840.95)(2.695,-0.447){3}{\rule{1.833pt}{0.108pt}}
\multiput(324.00,841.17)(9.195,-3.000){2}{\rule{0.917pt}{0.400pt}}
\multiput(337.00,837.95)(2.695,-0.447){3}{\rule{1.833pt}{0.108pt}}
\multiput(337.00,838.17)(9.195,-3.000){2}{\rule{0.917pt}{0.400pt}}
\multiput(350.00,834.95)(2.695,-0.447){3}{\rule{1.833pt}{0.108pt}}
\multiput(350.00,835.17)(9.195,-3.000){2}{\rule{0.917pt}{0.400pt}}
\multiput(363.00,831.95)(2.695,-0.447){3}{\rule{1.833pt}{0.108pt}}
\multiput(363.00,832.17)(9.195,-3.000){2}{\rule{0.917pt}{0.400pt}}
\multiput(376.00,828.95)(2.695,-0.447){3}{\rule{1.833pt}{0.108pt}}
\multiput(376.00,829.17)(9.195,-3.000){2}{\rule{0.917pt}{0.400pt}}
\multiput(389.00,825.94)(1.797,-0.468){5}{\rule{1.400pt}{0.113pt}}
\multiput(389.00,826.17)(10.094,-4.000){2}{\rule{0.700pt}{0.400pt}}
\multiput(402.00,821.95)(2.918,-0.447){3}{\rule{1.967pt}{0.108pt}}
\multiput(402.00,822.17)(9.918,-3.000){2}{\rule{0.983pt}{0.400pt}}
\multiput(416.00,818.94)(1.797,-0.468){5}{\rule{1.400pt}{0.113pt}}
\multiput(416.00,819.17)(10.094,-4.000){2}{\rule{0.700pt}{0.400pt}}
\multiput(429.00,814.94)(1.797,-0.468){5}{\rule{1.400pt}{0.113pt}}
\multiput(429.00,815.17)(10.094,-4.000){2}{\rule{0.700pt}{0.400pt}}
\multiput(442.00,810.93)(1.378,-0.477){7}{\rule{1.140pt}{0.115pt}}
\multiput(442.00,811.17)(10.634,-5.000){2}{\rule{0.570pt}{0.400pt}}
\multiput(455.00,805.94)(1.797,-0.468){5}{\rule{1.400pt}{0.113pt}}
\multiput(455.00,806.17)(10.094,-4.000){2}{\rule{0.700pt}{0.400pt}}
\multiput(468.00,801.93)(1.378,-0.477){7}{\rule{1.140pt}{0.115pt}}
\multiput(468.00,802.17)(10.634,-5.000){2}{\rule{0.570pt}{0.400pt}}
\multiput(481.00,796.94)(1.797,-0.468){5}{\rule{1.400pt}{0.113pt}}
\multiput(481.00,797.17)(10.094,-4.000){2}{\rule{0.700pt}{0.400pt}}
\multiput(494.00,792.93)(1.378,-0.477){7}{\rule{1.140pt}{0.115pt}}
\multiput(494.00,793.17)(10.634,-5.000){2}{\rule{0.570pt}{0.400pt}}
\multiput(507.00,787.93)(1.489,-0.477){7}{\rule{1.220pt}{0.115pt}}
\multiput(507.00,788.17)(11.468,-5.000){2}{\rule{0.610pt}{0.400pt}}
\multiput(521.00,782.93)(1.123,-0.482){9}{\rule{0.967pt}{0.116pt}}
\multiput(521.00,783.17)(10.994,-6.000){2}{\rule{0.483pt}{0.400pt}}
\multiput(534.00,776.93)(1.378,-0.477){7}{\rule{1.140pt}{0.115pt}}
\multiput(534.00,777.17)(10.634,-5.000){2}{\rule{0.570pt}{0.400pt}}
\multiput(547.00,771.93)(1.123,-0.482){9}{\rule{0.967pt}{0.116pt}}
\multiput(547.00,772.17)(10.994,-6.000){2}{\rule{0.483pt}{0.400pt}}
\multiput(560.00,765.93)(1.378,-0.477){7}{\rule{1.140pt}{0.115pt}}
\multiput(560.00,766.17)(10.634,-5.000){2}{\rule{0.570pt}{0.400pt}}
\multiput(573.00,760.93)(1.123,-0.482){9}{\rule{0.967pt}{0.116pt}}
\multiput(573.00,761.17)(10.994,-6.000){2}{\rule{0.483pt}{0.400pt}}
\multiput(586.00,754.93)(1.123,-0.482){9}{\rule{0.967pt}{0.116pt}}
\multiput(586.00,755.17)(10.994,-6.000){2}{\rule{0.483pt}{0.400pt}}
\multiput(599.00,748.93)(1.123,-0.482){9}{\rule{0.967pt}{0.116pt}}
\multiput(599.00,749.17)(10.994,-6.000){2}{\rule{0.483pt}{0.400pt}}
\multiput(612.00,742.93)(0.950,-0.485){11}{\rule{0.843pt}{0.117pt}}
\multiput(612.00,743.17)(11.251,-7.000){2}{\rule{0.421pt}{0.400pt}}
\multiput(625.00,735.93)(1.214,-0.482){9}{\rule{1.033pt}{0.116pt}}
\multiput(625.00,736.17)(11.855,-6.000){2}{\rule{0.517pt}{0.400pt}}
\multiput(639.00,729.93)(0.950,-0.485){11}{\rule{0.843pt}{0.117pt}}
\multiput(639.00,730.17)(11.251,-7.000){2}{\rule{0.421pt}{0.400pt}}
\multiput(652.00,722.93)(0.950,-0.485){11}{\rule{0.843pt}{0.117pt}}
\multiput(652.00,723.17)(11.251,-7.000){2}{\rule{0.421pt}{0.400pt}}
\multiput(665.00,715.93)(0.950,-0.485){11}{\rule{0.843pt}{0.117pt}}
\multiput(665.00,716.17)(11.251,-7.000){2}{\rule{0.421pt}{0.400pt}}
\multiput(678.00,708.93)(0.950,-0.485){11}{\rule{0.843pt}{0.117pt}}
\multiput(678.00,709.17)(11.251,-7.000){2}{\rule{0.421pt}{0.400pt}}
\multiput(691.00,701.93)(0.950,-0.485){11}{\rule{0.843pt}{0.117pt}}
\multiput(691.00,702.17)(11.251,-7.000){2}{\rule{0.421pt}{0.400pt}}
\multiput(704.00,694.93)(0.824,-0.488){13}{\rule{0.750pt}{0.117pt}}
\multiput(704.00,695.17)(11.443,-8.000){2}{\rule{0.375pt}{0.400pt}}
\multiput(717.00,686.93)(0.950,-0.485){11}{\rule{0.843pt}{0.117pt}}
\multiput(717.00,687.17)(11.251,-7.000){2}{\rule{0.421pt}{0.400pt}}
\multiput(730.00,679.93)(0.890,-0.488){13}{\rule{0.800pt}{0.117pt}}
\multiput(730.00,680.17)(12.340,-8.000){2}{\rule{0.400pt}{0.400pt}}
\multiput(744.00,671.93)(0.824,-0.488){13}{\rule{0.750pt}{0.117pt}}
\multiput(744.00,672.17)(11.443,-8.000){2}{\rule{0.375pt}{0.400pt}}
\multiput(757.00,663.93)(0.824,-0.488){13}{\rule{0.750pt}{0.117pt}}
\multiput(757.00,664.17)(11.443,-8.000){2}{\rule{0.375pt}{0.400pt}}
\multiput(770.00,655.93)(0.824,-0.488){13}{\rule{0.750pt}{0.117pt}}
\multiput(770.00,656.17)(11.443,-8.000){2}{\rule{0.375pt}{0.400pt}}
\multiput(783.00,647.93)(0.824,-0.488){13}{\rule{0.750pt}{0.117pt}}
\multiput(783.00,648.17)(11.443,-8.000){2}{\rule{0.375pt}{0.400pt}}
\multiput(796.00,639.93)(0.728,-0.489){15}{\rule{0.678pt}{0.118pt}}
\multiput(796.00,640.17)(11.593,-9.000){2}{\rule{0.339pt}{0.400pt}}
\multiput(809.00,630.93)(0.824,-0.488){13}{\rule{0.750pt}{0.117pt}}
\multiput(809.00,631.17)(11.443,-8.000){2}{\rule{0.375pt}{0.400pt}}
\multiput(822.00,622.93)(0.728,-0.489){15}{\rule{0.678pt}{0.118pt}}
\multiput(822.00,623.17)(11.593,-9.000){2}{\rule{0.339pt}{0.400pt}}
\multiput(835.00,613.93)(0.786,-0.489){15}{\rule{0.722pt}{0.118pt}}
\multiput(835.00,614.17)(12.501,-9.000){2}{\rule{0.361pt}{0.400pt}}
\multiput(849.00,604.93)(0.728,-0.489){15}{\rule{0.678pt}{0.118pt}}
\multiput(849.00,605.17)(11.593,-9.000){2}{\rule{0.339pt}{0.400pt}}
\multiput(862.00,595.93)(0.728,-0.489){15}{\rule{0.678pt}{0.118pt}}
\multiput(862.00,596.17)(11.593,-9.000){2}{\rule{0.339pt}{0.400pt}}
\multiput(875.00,586.93)(0.728,-0.489){15}{\rule{0.678pt}{0.118pt}}
\multiput(875.00,587.17)(11.593,-9.000){2}{\rule{0.339pt}{0.400pt}}
\multiput(888.00,577.93)(0.728,-0.489){15}{\rule{0.678pt}{0.118pt}}
\multiput(888.00,578.17)(11.593,-9.000){2}{\rule{0.339pt}{0.400pt}}
\multiput(901.00,568.92)(0.652,-0.491){17}{\rule{0.620pt}{0.118pt}}
\multiput(901.00,569.17)(11.713,-10.000){2}{\rule{0.310pt}{0.400pt}}
\multiput(914.00,558.93)(0.728,-0.489){15}{\rule{0.678pt}{0.118pt}}
\multiput(914.00,559.17)(11.593,-9.000){2}{\rule{0.339pt}{0.400pt}}
\multiput(927.00,549.92)(0.652,-0.491){17}{\rule{0.620pt}{0.118pt}}
\multiput(927.00,550.17)(11.713,-10.000){2}{\rule{0.310pt}{0.400pt}}
\multiput(940.00,539.92)(0.704,-0.491){17}{\rule{0.660pt}{0.118pt}}
\multiput(940.00,540.17)(12.630,-10.000){2}{\rule{0.330pt}{0.400pt}}
\multiput(954.00,529.92)(0.652,-0.491){17}{\rule{0.620pt}{0.118pt}}
\multiput(954.00,530.17)(11.713,-10.000){2}{\rule{0.310pt}{0.400pt}}
\multiput(967.00,519.92)(0.652,-0.491){17}{\rule{0.620pt}{0.118pt}}
\multiput(967.00,520.17)(11.713,-10.000){2}{\rule{0.310pt}{0.400pt}}
\multiput(980.00,509.92)(0.652,-0.491){17}{\rule{0.620pt}{0.118pt}}
\multiput(980.00,510.17)(11.713,-10.000){2}{\rule{0.310pt}{0.400pt}}
\multiput(993.00,499.92)(0.652,-0.491){17}{\rule{0.620pt}{0.118pt}}
\multiput(993.00,500.17)(11.713,-10.000){2}{\rule{0.310pt}{0.400pt}}
\multiput(1006.00,489.92)(0.590,-0.492){19}{\rule{0.573pt}{0.118pt}}
\multiput(1006.00,490.17)(11.811,-11.000){2}{\rule{0.286pt}{0.400pt}}
\multiput(1019.00,478.92)(0.652,-0.491){17}{\rule{0.620pt}{0.118pt}}
\multiput(1019.00,479.17)(11.713,-10.000){2}{\rule{0.310pt}{0.400pt}}
\multiput(1032.00,468.92)(0.590,-0.492){19}{\rule{0.573pt}{0.118pt}}
\multiput(1032.00,469.17)(11.811,-11.000){2}{\rule{0.286pt}{0.400pt}}
\multiput(1045.00,457.92)(0.590,-0.492){19}{\rule{0.573pt}{0.118pt}}
\multiput(1045.00,458.17)(11.811,-11.000){2}{\rule{0.286pt}{0.400pt}}
\multiput(1058.00,446.92)(0.637,-0.492){19}{\rule{0.609pt}{0.118pt}}
\multiput(1058.00,447.17)(12.736,-11.000){2}{\rule{0.305pt}{0.400pt}}
\multiput(1072.00,435.92)(0.590,-0.492){19}{\rule{0.573pt}{0.118pt}}
\multiput(1072.00,436.17)(11.811,-11.000){2}{\rule{0.286pt}{0.400pt}}
\multiput(1085.00,424.92)(0.590,-0.492){19}{\rule{0.573pt}{0.118pt}}
\multiput(1085.00,425.17)(11.811,-11.000){2}{\rule{0.286pt}{0.400pt}}
\multiput(1098.00,413.92)(0.590,-0.492){19}{\rule{0.573pt}{0.118pt}}
\multiput(1098.00,414.17)(11.811,-11.000){2}{\rule{0.286pt}{0.400pt}}
\multiput(1111.00,402.92)(0.590,-0.492){19}{\rule{0.573pt}{0.118pt}}
\multiput(1111.00,403.17)(11.811,-11.000){2}{\rule{0.286pt}{0.400pt}}
\multiput(1124.00,391.92)(0.590,-0.492){19}{\rule{0.573pt}{0.118pt}}
\multiput(1124.00,392.17)(11.811,-11.000){2}{\rule{0.286pt}{0.400pt}}
\multiput(1137.00,380.92)(0.539,-0.492){21}{\rule{0.533pt}{0.119pt}}
\multiput(1137.00,381.17)(11.893,-12.000){2}{\rule{0.267pt}{0.400pt}}
\multiput(1150.00,368.92)(0.590,-0.492){19}{\rule{0.573pt}{0.118pt}}
\multiput(1150.00,369.17)(11.811,-11.000){2}{\rule{0.286pt}{0.400pt}}
\multiput(1163.00,357.92)(0.582,-0.492){21}{\rule{0.567pt}{0.119pt}}
\multiput(1163.00,358.17)(12.824,-12.000){2}{\rule{0.283pt}{0.400pt}}
\multiput(1177.00,345.92)(0.590,-0.492){19}{\rule{0.573pt}{0.118pt}}
\multiput(1177.00,346.17)(11.811,-11.000){2}{\rule{0.286pt}{0.400pt}}
\multiput(1190.00,334.92)(0.539,-0.492){21}{\rule{0.533pt}{0.119pt}}
\multiput(1190.00,335.17)(11.893,-12.000){2}{\rule{0.267pt}{0.400pt}}
\multiput(1203.00,322.92)(0.539,-0.492){21}{\rule{0.533pt}{0.119pt}}
\multiput(1203.00,323.17)(11.893,-12.000){2}{\rule{0.267pt}{0.400pt}}
\multiput(1216.00,310.92)(0.539,-0.492){21}{\rule{0.533pt}{0.119pt}}
\multiput(1216.00,311.17)(11.893,-12.000){2}{\rule{0.267pt}{0.400pt}}
\multiput(1229.00,298.92)(0.539,-0.492){21}{\rule{0.533pt}{0.119pt}}
\multiput(1229.00,299.17)(11.893,-12.000){2}{\rule{0.267pt}{0.400pt}}
\multiput(1242.00,286.92)(0.539,-0.492){21}{\rule{0.533pt}{0.119pt}}
\multiput(1242.00,287.17)(11.893,-12.000){2}{\rule{0.267pt}{0.400pt}}
\multiput(1255.00,274.92)(0.539,-0.492){21}{\rule{0.533pt}{0.119pt}}
\multiput(1255.00,275.17)(11.893,-12.000){2}{\rule{0.267pt}{0.400pt}}
\multiput(1268.00,262.92)(0.536,-0.493){23}{\rule{0.531pt}{0.119pt}}
\multiput(1268.00,263.17)(12.898,-13.000){2}{\rule{0.265pt}{0.400pt}}
\multiput(1282.00,249.92)(0.539,-0.492){21}{\rule{0.533pt}{0.119pt}}
\multiput(1282.00,250.17)(11.893,-12.000){2}{\rule{0.267pt}{0.400pt}}
\multiput(1295.00,237.92)(0.539,-0.492){21}{\rule{0.533pt}{0.119pt}}
\multiput(1295.00,238.17)(11.893,-12.000){2}{\rule{0.267pt}{0.400pt}}
\multiput(1308.00,225.92)(0.497,-0.493){23}{\rule{0.500pt}{0.119pt}}
\multiput(1308.00,226.17)(11.962,-13.000){2}{\rule{0.250pt}{0.400pt}}
\multiput(1321.00,212.92)(0.539,-0.492){21}{\rule{0.533pt}{0.119pt}}
\multiput(1321.00,213.17)(11.893,-12.000){2}{\rule{0.267pt}{0.400pt}}
\multiput(1334.00,200.92)(0.497,-0.493){23}{\rule{0.500pt}{0.119pt}}
\multiput(1334.00,201.17)(11.962,-13.000){2}{\rule{0.250pt}{0.400pt}}
\multiput(1347.00,187.92)(0.497,-0.493){23}{\rule{0.500pt}{0.119pt}}
\multiput(1347.00,188.17)(11.962,-13.000){2}{\rule{0.250pt}{0.400pt}}
\multiput(1360.00,174.92)(0.497,-0.493){23}{\rule{0.500pt}{0.119pt}}
\multiput(1360.00,175.17)(11.962,-13.000){2}{\rule{0.250pt}{0.400pt}}
\multiput(1373.00,161.92)(0.582,-0.492){21}{\rule{0.567pt}{0.119pt}}
\multiput(1373.00,162.17)(12.824,-12.000){2}{\rule{0.283pt}{0.400pt}}
\multiput(1387.00,149.92)(0.497,-0.493){23}{\rule{0.500pt}{0.119pt}}
\multiput(1387.00,150.17)(11.962,-13.000){2}{\rule{0.250pt}{0.400pt}}
\multiput(1400.00,136.92)(0.497,-0.493){23}{\rule{0.500pt}{0.119pt}}
\multiput(1400.00,137.17)(11.962,-13.000){2}{\rule{0.250pt}{0.400pt}}
\multiput(1413.00,123.92)(0.497,-0.493){23}{\rule{0.500pt}{0.119pt}}
\multiput(1413.00,124.17)(11.962,-13.000){2}{\rule{0.250pt}{0.400pt}}
\multiput(1426.00,110.92)(0.497,-0.493){23}{\rule{0.500pt}{0.119pt}}
\multiput(1426.00,111.17)(11.962,-13.000){2}{\rule{0.250pt}{0.400pt}}
\put(179.0,859.0){\rule[-0.200pt]{3.132pt}{0.400pt}}

\put(1399,779){\usebox{\plotpoint}}
\put(140,860){\usebox{\plotpoint}}
\put(140.00,860.00){\usebox{\plotpoint}}
\put(160.76,860.00){\usebox{\plotpoint}}
\put(181.47,859.00){\usebox{\plotpoint}}
\put(202.20,858.27){\usebox{\plotpoint}}
\put(222.91,857.00){\usebox{\plotpoint}}
\put(243.63,856.11){\usebox{\plotpoint}}
\put(264.21,853.52){\usebox{\plotpoint}}
\put(284.90,851.86){\usebox{\plotpoint}}
\put(305.42,848.80){\usebox{\plotpoint}}
\put(325.97,845.85){\usebox{\plotpoint}}
\put(346.44,842.82){\usebox{\plotpoint}}
\put(366.90,839.40){\usebox{\plotpoint}}
\put(387.26,835.40){\usebox{\plotpoint}}
\put(407.68,831.78){\usebox{\plotpoint}}
\put(427.94,827.25){\usebox{\plotpoint}}
\put(448.16,822.58){\usebox{\plotpoint}}
\put(468.38,817.91){\usebox{\plotpoint}}
\put(488.46,812.70){\usebox{\plotpoint}}
\put(508.56,807.55){\usebox{\plotpoint}}
\put(528.47,801.70){\usebox{\plotpoint}}
\put(548.56,796.52){\usebox{\plotpoint}}
\put(568.20,789.85){\usebox{\plotpoint}}
\put(587.92,783.41){\usebox{\plotpoint}}
\put(607.55,776.71){\usebox{\plotpoint}}
\put(627.25,770.20){\usebox{\plotpoint}}
\put(646.91,763.57){\usebox{\plotpoint}}
\put(666.40,756.46){\usebox{\plotpoint}}
\put(685.77,749.01){\usebox{\plotpoint}}
\put(705.12,741.49){\usebox{\plotpoint}}
\put(724.16,733.25){\usebox{\plotpoint}}
\put(743.65,726.13){\usebox{\plotpoint}}
\put(762.51,717.46){\usebox{\plotpoint}}
\put(781.67,709.51){\usebox{\plotpoint}}
\put(800.55,700.90){\usebox{\plotpoint}}
\put(819.40,692.20){\usebox{\plotpoint}}
\put(838.28,683.59){\usebox{\plotpoint}}
\put(857.01,674.69){\usebox{\plotpoint}}
\put(875.70,665.68){\usebox{\plotpoint}}
\put(894.34,656.58){\usebox{\plotpoint}}
\put(912.62,646.74){\usebox{\plotpoint}}
\put(931.29,637.69){\usebox{\plotpoint}}
\put(949.71,628.14){\usebox{\plotpoint}}
\put(968.05,618.43){\usebox{\plotpoint}}
\put(986.33,608.59){\usebox{\plotpoint}}
\put(1004.60,598.75){\usebox{\plotpoint}}
\put(1022.44,588.15){\usebox{\plotpoint}}
\put(1040.71,578.31){\usebox{\plotpoint}}
\put(1058.54,567.69){\usebox{\plotpoint}}
\put(1076.62,557.51){\usebox{\plotpoint}}
\put(1094.58,547.11){\usebox{\plotpoint}}
\put(1112.25,536.23){\usebox{\plotpoint}}
\put(1129.93,525.35){\usebox{\plotpoint}}
\put(1147.61,514.47){\usebox{\plotpoint}}
\put(1165.33,503.67){\usebox{\plotpoint}}
\put(1183.23,493.17){\usebox{\plotpoint}}
\put(1200.90,482.29){\usebox{\plotpoint}}
\put(1218.11,470.70){\usebox{\plotpoint}}
\put(1235.55,459.46){\usebox{\plotpoint}}
\put(1253.00,448.23){\usebox{\plotpoint}}
\put(1270.25,436.71){\usebox{\plotpoint}}
\put(1287.94,425.89){\usebox{\plotpoint}}
\put(1305.01,414.07){\usebox{\plotpoint}}
\put(1322.07,402.26){\usebox{\plotpoint}}
\put(1339.14,390.44){\usebox{\plotpoint}}
\put(1356.20,378.63){\usebox{\plotpoint}}
\put(1373.27,366.82){\usebox{\plotpoint}}
\put(1390.65,355.47){\usebox{\plotpoint}}
\put(1407.71,343.66){\usebox{\plotpoint}}
\put(1424.78,331.85){\usebox{\plotpoint}}
\put(1439,321){\usebox{\plotpoint}}

\sbox{\plotpoint}{\rule[-0.400pt]{0.800pt}{0.800pt}}%
\put(140,860){\usebox{\plotpoint}}
\put(166,857.84){\rule{3.132pt}{0.800pt}}
\multiput(166.00,858.34)(6.500,-1.000){2}{\rule{1.566pt}{0.800pt}}
\put(179,856.84){\rule{3.132pt}{0.800pt}}
\multiput(179.00,857.34)(6.500,-1.000){2}{\rule{1.566pt}{0.800pt}}
\put(192,855.84){\rule{3.373pt}{0.800pt}}
\multiput(192.00,856.34)(7.000,-1.000){2}{\rule{1.686pt}{0.800pt}}
\put(206,854.84){\rule{3.132pt}{0.800pt}}
\multiput(206.00,855.34)(6.500,-1.000){2}{\rule{1.566pt}{0.800pt}}
\put(219,853.34){\rule{3.132pt}{0.800pt}}
\multiput(219.00,854.34)(6.500,-2.000){2}{\rule{1.566pt}{0.800pt}}
\put(232,851.84){\rule{3.132pt}{0.800pt}}
\multiput(232.00,852.34)(6.500,-1.000){2}{\rule{1.566pt}{0.800pt}}
\put(245,850.34){\rule{3.132pt}{0.800pt}}
\multiput(245.00,851.34)(6.500,-2.000){2}{\rule{1.566pt}{0.800pt}}
\put(258,848.34){\rule{3.132pt}{0.800pt}}
\multiput(258.00,849.34)(6.500,-2.000){2}{\rule{1.566pt}{0.800pt}}
\put(271,846.34){\rule{3.132pt}{0.800pt}}
\multiput(271.00,847.34)(6.500,-2.000){2}{\rule{1.566pt}{0.800pt}}
\put(284,843.84){\rule{3.132pt}{0.800pt}}
\multiput(284.00,845.34)(6.500,-3.000){2}{\rule{1.566pt}{0.800pt}}
\put(297,840.84){\rule{3.373pt}{0.800pt}}
\multiput(297.00,842.34)(7.000,-3.000){2}{\rule{1.686pt}{0.800pt}}
\put(311,838.34){\rule{3.132pt}{0.800pt}}
\multiput(311.00,839.34)(6.500,-2.000){2}{\rule{1.566pt}{0.800pt}}
\put(324,835.84){\rule{3.132pt}{0.800pt}}
\multiput(324.00,837.34)(6.500,-3.000){2}{\rule{1.566pt}{0.800pt}}
\put(337,832.34){\rule{2.800pt}{0.800pt}}
\multiput(337.00,834.34)(7.188,-4.000){2}{\rule{1.400pt}{0.800pt}}
\put(350,828.84){\rule{3.132pt}{0.800pt}}
\multiput(350.00,830.34)(6.500,-3.000){2}{\rule{1.566pt}{0.800pt}}
\put(363,825.34){\rule{2.800pt}{0.800pt}}
\multiput(363.00,827.34)(7.188,-4.000){2}{\rule{1.400pt}{0.800pt}}
\put(376,821.84){\rule{3.132pt}{0.800pt}}
\multiput(376.00,823.34)(6.500,-3.000){2}{\rule{1.566pt}{0.800pt}}
\put(389,818.34){\rule{2.800pt}{0.800pt}}
\multiput(389.00,820.34)(7.188,-4.000){2}{\rule{1.400pt}{0.800pt}}
\multiput(402.00,816.06)(1.936,-0.560){3}{\rule{2.440pt}{0.135pt}}
\multiput(402.00,816.34)(8.936,-5.000){2}{\rule{1.220pt}{0.800pt}}
\put(416,809.34){\rule{2.800pt}{0.800pt}}
\multiput(416.00,811.34)(7.188,-4.000){2}{\rule{1.400pt}{0.800pt}}
\put(429,805.34){\rule{2.800pt}{0.800pt}}
\multiput(429.00,807.34)(7.188,-4.000){2}{\rule{1.400pt}{0.800pt}}
\multiput(442.00,803.06)(1.768,-0.560){3}{\rule{2.280pt}{0.135pt}}
\multiput(442.00,803.34)(8.268,-5.000){2}{\rule{1.140pt}{0.800pt}}
\multiput(455.00,798.06)(1.768,-0.560){3}{\rule{2.280pt}{0.135pt}}
\multiput(455.00,798.34)(8.268,-5.000){2}{\rule{1.140pt}{0.800pt}}
\multiput(468.00,793.06)(1.768,-0.560){3}{\rule{2.280pt}{0.135pt}}
\multiput(468.00,793.34)(8.268,-5.000){2}{\rule{1.140pt}{0.800pt}}
\multiput(481.00,788.06)(1.768,-0.560){3}{\rule{2.280pt}{0.135pt}}
\multiput(481.00,788.34)(8.268,-5.000){2}{\rule{1.140pt}{0.800pt}}
\multiput(494.00,783.06)(1.768,-0.560){3}{\rule{2.280pt}{0.135pt}}
\multiput(494.00,783.34)(8.268,-5.000){2}{\rule{1.140pt}{0.800pt}}
\multiput(507.00,778.06)(1.936,-0.560){3}{\rule{2.440pt}{0.135pt}}
\multiput(507.00,778.34)(8.936,-5.000){2}{\rule{1.220pt}{0.800pt}}
\multiput(521.00,773.07)(1.244,-0.536){5}{\rule{1.933pt}{0.129pt}}
\multiput(521.00,773.34)(8.987,-6.000){2}{\rule{0.967pt}{0.800pt}}
\multiput(534.00,767.07)(1.244,-0.536){5}{\rule{1.933pt}{0.129pt}}
\multiput(534.00,767.34)(8.987,-6.000){2}{\rule{0.967pt}{0.800pt}}
\multiput(547.00,761.07)(1.244,-0.536){5}{\rule{1.933pt}{0.129pt}}
\multiput(547.00,761.34)(8.987,-6.000){2}{\rule{0.967pt}{0.800pt}}
\multiput(560.00,755.07)(1.244,-0.536){5}{\rule{1.933pt}{0.129pt}}
\multiput(560.00,755.34)(8.987,-6.000){2}{\rule{0.967pt}{0.800pt}}
\multiput(573.00,749.07)(1.244,-0.536){5}{\rule{1.933pt}{0.129pt}}
\multiput(573.00,749.34)(8.987,-6.000){2}{\rule{0.967pt}{0.800pt}}
\multiput(586.00,743.07)(1.244,-0.536){5}{\rule{1.933pt}{0.129pt}}
\multiput(586.00,743.34)(8.987,-6.000){2}{\rule{0.967pt}{0.800pt}}
\multiput(599.00,737.07)(1.244,-0.536){5}{\rule{1.933pt}{0.129pt}}
\multiput(599.00,737.34)(8.987,-6.000){2}{\rule{0.967pt}{0.800pt}}
\multiput(612.00,731.08)(1.000,-0.526){7}{\rule{1.686pt}{0.127pt}}
\multiput(612.00,731.34)(9.501,-7.000){2}{\rule{0.843pt}{0.800pt}}
\multiput(625.00,724.08)(1.088,-0.526){7}{\rule{1.800pt}{0.127pt}}
\multiput(625.00,724.34)(10.264,-7.000){2}{\rule{0.900pt}{0.800pt}}
\multiput(639.00,717.07)(1.244,-0.536){5}{\rule{1.933pt}{0.129pt}}
\multiput(639.00,717.34)(8.987,-6.000){2}{\rule{0.967pt}{0.800pt}}
\multiput(652.00,711.08)(1.000,-0.526){7}{\rule{1.686pt}{0.127pt}}
\multiput(652.00,711.34)(9.501,-7.000){2}{\rule{0.843pt}{0.800pt}}
\multiput(665.00,704.08)(1.000,-0.526){7}{\rule{1.686pt}{0.127pt}}
\multiput(665.00,704.34)(9.501,-7.000){2}{\rule{0.843pt}{0.800pt}}
\multiput(678.00,697.08)(1.000,-0.526){7}{\rule{1.686pt}{0.127pt}}
\multiput(678.00,697.34)(9.501,-7.000){2}{\rule{0.843pt}{0.800pt}}
\multiput(691.00,690.08)(0.847,-0.520){9}{\rule{1.500pt}{0.125pt}}
\multiput(691.00,690.34)(9.887,-8.000){2}{\rule{0.750pt}{0.800pt}}
\multiput(704.00,682.08)(1.000,-0.526){7}{\rule{1.686pt}{0.127pt}}
\multiput(704.00,682.34)(9.501,-7.000){2}{\rule{0.843pt}{0.800pt}}
\multiput(717.00,675.08)(0.847,-0.520){9}{\rule{1.500pt}{0.125pt}}
\multiput(717.00,675.34)(9.887,-8.000){2}{\rule{0.750pt}{0.800pt}}
\multiput(730.00,667.08)(1.088,-0.526){7}{\rule{1.800pt}{0.127pt}}
\multiput(730.00,667.34)(10.264,-7.000){2}{\rule{0.900pt}{0.800pt}}
\multiput(744.00,660.08)(0.847,-0.520){9}{\rule{1.500pt}{0.125pt}}
\multiput(744.00,660.34)(9.887,-8.000){2}{\rule{0.750pt}{0.800pt}}
\multiput(757.00,652.08)(0.847,-0.520){9}{\rule{1.500pt}{0.125pt}}
\multiput(757.00,652.34)(9.887,-8.000){2}{\rule{0.750pt}{0.800pt}}
\multiput(770.00,644.08)(0.847,-0.520){9}{\rule{1.500pt}{0.125pt}}
\multiput(770.00,644.34)(9.887,-8.000){2}{\rule{0.750pt}{0.800pt}}
\multiput(783.00,636.08)(0.847,-0.520){9}{\rule{1.500pt}{0.125pt}}
\multiput(783.00,636.34)(9.887,-8.000){2}{\rule{0.750pt}{0.800pt}}
\multiput(796.00,628.08)(0.847,-0.520){9}{\rule{1.500pt}{0.125pt}}
\multiput(796.00,628.34)(9.887,-8.000){2}{\rule{0.750pt}{0.800pt}}
\multiput(809.00,620.08)(0.847,-0.520){9}{\rule{1.500pt}{0.125pt}}
\multiput(809.00,620.34)(9.887,-8.000){2}{\rule{0.750pt}{0.800pt}}
\multiput(822.00,612.08)(0.737,-0.516){11}{\rule{1.356pt}{0.124pt}}
\multiput(822.00,612.34)(10.186,-9.000){2}{\rule{0.678pt}{0.800pt}}
\multiput(835.00,603.08)(0.920,-0.520){9}{\rule{1.600pt}{0.125pt}}
\multiput(835.00,603.34)(10.679,-8.000){2}{\rule{0.800pt}{0.800pt}}
\multiput(849.00,595.08)(0.737,-0.516){11}{\rule{1.356pt}{0.124pt}}
\multiput(849.00,595.34)(10.186,-9.000){2}{\rule{0.678pt}{0.800pt}}
\multiput(862.00,586.08)(0.847,-0.520){9}{\rule{1.500pt}{0.125pt}}
\multiput(862.00,586.34)(9.887,-8.000){2}{\rule{0.750pt}{0.800pt}}
\multiput(875.00,578.08)(0.737,-0.516){11}{\rule{1.356pt}{0.124pt}}
\multiput(875.00,578.34)(10.186,-9.000){2}{\rule{0.678pt}{0.800pt}}
\multiput(888.00,569.08)(0.737,-0.516){11}{\rule{1.356pt}{0.124pt}}
\multiput(888.00,569.34)(10.186,-9.000){2}{\rule{0.678pt}{0.800pt}}
\multiput(901.00,560.08)(0.737,-0.516){11}{\rule{1.356pt}{0.124pt}}
\multiput(901.00,560.34)(10.186,-9.000){2}{\rule{0.678pt}{0.800pt}}
\multiput(914.00,551.08)(0.737,-0.516){11}{\rule{1.356pt}{0.124pt}}
\multiput(914.00,551.34)(10.186,-9.000){2}{\rule{0.678pt}{0.800pt}}
\multiput(927.00,542.08)(0.737,-0.516){11}{\rule{1.356pt}{0.124pt}}
\multiput(927.00,542.34)(10.186,-9.000){2}{\rule{0.678pt}{0.800pt}}
\multiput(940.00,533.08)(0.800,-0.516){11}{\rule{1.444pt}{0.124pt}}
\multiput(940.00,533.34)(11.002,-9.000){2}{\rule{0.722pt}{0.800pt}}
\multiput(954.00,524.08)(0.737,-0.516){11}{\rule{1.356pt}{0.124pt}}
\multiput(954.00,524.34)(10.186,-9.000){2}{\rule{0.678pt}{0.800pt}}
\multiput(967.00,515.08)(0.654,-0.514){13}{\rule{1.240pt}{0.124pt}}
\multiput(967.00,515.34)(10.426,-10.000){2}{\rule{0.620pt}{0.800pt}}
\multiput(980.00,505.08)(0.737,-0.516){11}{\rule{1.356pt}{0.124pt}}
\multiput(980.00,505.34)(10.186,-9.000){2}{\rule{0.678pt}{0.800pt}}
\multiput(993.00,496.08)(0.654,-0.514){13}{\rule{1.240pt}{0.124pt}}
\multiput(993.00,496.34)(10.426,-10.000){2}{\rule{0.620pt}{0.800pt}}
\multiput(1006.00,486.08)(0.737,-0.516){11}{\rule{1.356pt}{0.124pt}}
\multiput(1006.00,486.34)(10.186,-9.000){2}{\rule{0.678pt}{0.800pt}}
\multiput(1019.00,477.08)(0.654,-0.514){13}{\rule{1.240pt}{0.124pt}}
\multiput(1019.00,477.34)(10.426,-10.000){2}{\rule{0.620pt}{0.800pt}}
\multiput(1032.00,467.08)(0.737,-0.516){11}{\rule{1.356pt}{0.124pt}}
\multiput(1032.00,467.34)(10.186,-9.000){2}{\rule{0.678pt}{0.800pt}}
\multiput(1045.00,458.08)(0.654,-0.514){13}{\rule{1.240pt}{0.124pt}}
\multiput(1045.00,458.34)(10.426,-10.000){2}{\rule{0.620pt}{0.800pt}}
\multiput(1058.00,448.08)(0.710,-0.514){13}{\rule{1.320pt}{0.124pt}}
\multiput(1058.00,448.34)(11.260,-10.000){2}{\rule{0.660pt}{0.800pt}}
\multiput(1072.00,438.08)(0.654,-0.514){13}{\rule{1.240pt}{0.124pt}}
\multiput(1072.00,438.34)(10.426,-10.000){2}{\rule{0.620pt}{0.800pt}}
\multiput(1085.00,428.08)(0.654,-0.514){13}{\rule{1.240pt}{0.124pt}}
\multiput(1085.00,428.34)(10.426,-10.000){2}{\rule{0.620pt}{0.800pt}}
\multiput(1098.00,418.08)(0.654,-0.514){13}{\rule{1.240pt}{0.124pt}}
\multiput(1098.00,418.34)(10.426,-10.000){2}{\rule{0.620pt}{0.800pt}}
\multiput(1111.00,408.08)(0.654,-0.514){13}{\rule{1.240pt}{0.124pt}}
\multiput(1111.00,408.34)(10.426,-10.000){2}{\rule{0.620pt}{0.800pt}}
\multiput(1124.00,398.08)(0.654,-0.514){13}{\rule{1.240pt}{0.124pt}}
\multiput(1124.00,398.34)(10.426,-10.000){2}{\rule{0.620pt}{0.800pt}}
\multiput(1137.00,388.08)(0.654,-0.514){13}{\rule{1.240pt}{0.124pt}}
\multiput(1137.00,388.34)(10.426,-10.000){2}{\rule{0.620pt}{0.800pt}}
\multiput(1150.00,378.08)(0.589,-0.512){15}{\rule{1.145pt}{0.123pt}}
\multiput(1150.00,378.34)(10.623,-11.000){2}{\rule{0.573pt}{0.800pt}}
\multiput(1163.00,367.08)(0.710,-0.514){13}{\rule{1.320pt}{0.124pt}}
\multiput(1163.00,367.34)(11.260,-10.000){2}{\rule{0.660pt}{0.800pt}}
\multiput(1177.00,357.08)(0.654,-0.514){13}{\rule{1.240pt}{0.124pt}}
\multiput(1177.00,357.34)(10.426,-10.000){2}{\rule{0.620pt}{0.800pt}}
\multiput(1190.00,347.08)(0.589,-0.512){15}{\rule{1.145pt}{0.123pt}}
\multiput(1190.00,347.34)(10.623,-11.000){2}{\rule{0.573pt}{0.800pt}}
\multiput(1203.00,336.08)(0.654,-0.514){13}{\rule{1.240pt}{0.124pt}}
\multiput(1203.00,336.34)(10.426,-10.000){2}{\rule{0.620pt}{0.800pt}}
\multiput(1216.00,326.08)(0.589,-0.512){15}{\rule{1.145pt}{0.123pt}}
\multiput(1216.00,326.34)(10.623,-11.000){2}{\rule{0.573pt}{0.800pt}}
\multiput(1229.00,315.08)(0.654,-0.514){13}{\rule{1.240pt}{0.124pt}}
\multiput(1229.00,315.34)(10.426,-10.000){2}{\rule{0.620pt}{0.800pt}}
\multiput(1242.00,305.08)(0.589,-0.512){15}{\rule{1.145pt}{0.123pt}}
\multiput(1242.00,305.34)(10.623,-11.000){2}{\rule{0.573pt}{0.800pt}}
\multiput(1255.00,294.08)(0.589,-0.512){15}{\rule{1.145pt}{0.123pt}}
\multiput(1255.00,294.34)(10.623,-11.000){2}{\rule{0.573pt}{0.800pt}}
\multiput(1268.00,283.08)(0.710,-0.514){13}{\rule{1.320pt}{0.124pt}}
\multiput(1268.00,283.34)(11.260,-10.000){2}{\rule{0.660pt}{0.800pt}}
\multiput(1282.00,273.08)(0.589,-0.512){15}{\rule{1.145pt}{0.123pt}}
\multiput(1282.00,273.34)(10.623,-11.000){2}{\rule{0.573pt}{0.800pt}}
\multiput(1295.00,262.08)(0.589,-0.512){15}{\rule{1.145pt}{0.123pt}}
\multiput(1295.00,262.34)(10.623,-11.000){2}{\rule{0.573pt}{0.800pt}}
\multiput(1308.00,251.08)(0.589,-0.512){15}{\rule{1.145pt}{0.123pt}}
\multiput(1308.00,251.34)(10.623,-11.000){2}{\rule{0.573pt}{0.800pt}}
\multiput(1321.00,240.08)(0.589,-0.512){15}{\rule{1.145pt}{0.123pt}}
\multiput(1321.00,240.34)(10.623,-11.000){2}{\rule{0.573pt}{0.800pt}}
\multiput(1334.00,229.08)(0.654,-0.514){13}{\rule{1.240pt}{0.124pt}}
\multiput(1334.00,229.34)(10.426,-10.000){2}{\rule{0.620pt}{0.800pt}}
\multiput(1347.00,219.08)(0.589,-0.512){15}{\rule{1.145pt}{0.123pt}}
\multiput(1347.00,219.34)(10.623,-11.000){2}{\rule{0.573pt}{0.800pt}}
\multiput(1360.00,208.08)(0.589,-0.512){15}{\rule{1.145pt}{0.123pt}}
\multiput(1360.00,208.34)(10.623,-11.000){2}{\rule{0.573pt}{0.800pt}}
\multiput(1373.00,197.08)(0.639,-0.512){15}{\rule{1.218pt}{0.123pt}}
\multiput(1373.00,197.34)(11.472,-11.000){2}{\rule{0.609pt}{0.800pt}}
\multiput(1387.00,186.08)(0.536,-0.511){17}{\rule{1.067pt}{0.123pt}}
\multiput(1387.00,186.34)(10.786,-12.000){2}{\rule{0.533pt}{0.800pt}}
\multiput(1400.00,174.08)(0.589,-0.512){15}{\rule{1.145pt}{0.123pt}}
\multiput(1400.00,174.34)(10.623,-11.000){2}{\rule{0.573pt}{0.800pt}}
\multiput(1413.00,163.08)(0.589,-0.512){15}{\rule{1.145pt}{0.123pt}}
\multiput(1413.00,163.34)(10.623,-11.000){2}{\rule{0.573pt}{0.800pt}}
\multiput(1426.00,152.08)(0.589,-0.512){15}{\rule{1.145pt}{0.123pt}}
\multiput(1426.00,152.34)(10.623,-11.000){2}{\rule{0.573pt}{0.800pt}}
\put(140.0,860.0){\rule[-0.400pt]{6.263pt}{0.800pt}}

\put(1279,760){\makebox(0,0)[r]{$h_{S_7}(p)$}}
\put(1299.0,760){\rule[-0.400pt]{24.090pt}{0.800pt}}
\put(1279,700){\makebox(0,0)[r]{$h_{S_9}(p)$}}
\put(1299.0,700){\rule[-0.200pt]{24.090pt}{0.400pt}}
\put(1279,820){\makebox(0,0)[r]{$h_{S_5}(p)$}}
\multiput(1299,820)(20.756,0.000){5}{\usebox{\plotpoint}}

\end{picture}
\end{center}
\bigskip

As one can see, the effectiveness of these codes has an interesting
dependence on $p$ when errors on multiple qubits are figured in.  Comparing the Shor and Steane codes, for example, the
curves $h_{S_9}(p)$ and $h_{S_7}(p)$ cross near $p \approx .138$.

\section{Syndrome quality}\label{syndromesection}
Presumably, one will not know which qubits, or even how many, have undergone an error within a quantum computer.  The only accessible information about possible errors will be the measured syndrome, and not all error syndromes should be treated equally.  For example, in the Shor code, if the error syndrome $(01010111)$ is measured, there probability that the error correction step will succeed is zero.  Define the \emph{syndrome quality} of $\e\in \F_2^r$ to be the probability that error correction will succeed given that the error syndrome $\e$ is measured.  An actual implementation of a code should contain classical subroutines between the detection/correction phases to abort if a particularly low syndrome quality is encountered.  It is certainly better to know that qubits should be discarded, than it is to receive junk and believe it to be pristine.

Consider the probability that the code
 will correct an error affecting $t$ qubits if the syndrome $\e$ is measured.   One simply counts, for a given $\e$, the number of sets $\{m\}$ with $|\{m\}|=t$ for which $\Gm_\e$ satisfies condition (\ref{condition1}). 
It would be difficult to summarize this information in a table since there are $2^r$ distinct $\e$.  However, for the three codes discussed, the values vary only as $\e$ varies among a small number of sets.  We give the results.

In the table below, the first number in each column is the probability that the $[[5,1,3]]$ code will correct an error affecting $t$ qubits after the syndrome $\e$ is measured.  The second number is the probability that if an error affects $t$ qubits, that the syndrome $\e$ will be measured.

\begin{equation}
\begin{array}{c|cc|cc|cc|cc|cc|cc}
\text{syndrome} &\multicolumn{2}{|c}{t=0} &\multicolumn{2}{|c}{t=1} &\multicolumn{2}{|c}{t=2} &\multicolumn{2}{|c}{t=3} &\multicolumn{2}{|c}{t=4} &\multicolumn{2}{|c}{t=5} \\
\hline 
\e=(0000) & 1 & 1 & 1 & \frac{1}{4} & 1 & \frac{1}{16} & 0 & \frac{1}{16} & 0 & \frac{1}{16} & 0 & \frac{1}{16} \\ 
\e\neq(0000) & n/a & 0 & 1 & \frac{1}{20} & \frac{2}{5} & \frac{1}{16} & 0 & \frac{1}{16} & 0 & \frac{1}{16} & 0 & \frac{1}{16} \\ 
\end{array}
\end{equation}

Now let us consider the $[[7,3,1]]$ code.  The sixty-four distinct vectors $\e\in \F_2^6$ fall into the following three sets:
\begin{align}
A&=\{(000000)\},\\
B&=\{(000001),(000010),(000100),(000011),(000101),(000110),(000111),\notag \\
&(001001),(010010),(100100),(101101),(011011),(110110),(111111),\\
&(100000),(010000),(001000),(110000),(101000),(011000),(111000)\},\notag \\
C&=\F_2^6 \setminus (A \cup B).
\end{align}

Again, the first number in each column is the probability that the $[[5,1,3]]$ code will correct an error affecting $t$ qubits after the syndrome $\e$ is measured.  The second number is the probability that if an error affects $t$ qubits, that the syndrome $\e$ will be measured.

\begin{equation}
\begin{array}{c|cc|cc|cc|cc|cc|cc|cc|cc}
\text{syndrome} &\multicolumn{2}{|c}{t=0} &\multicolumn{2}{|c}{t=1} &\multicolumn{2}{|c}{t=2} &\multicolumn{2}{|c}{t=3} &\multicolumn{2}{|c}{t=4} &\multicolumn{2}{|c}{t=5} &\multicolumn{2}{|c}{t=6} &\multicolumn{2}{|c}{t=7} \\
\hline 
\e\in A & 1 & 1 & 1 & \frac{1}{4} & 1 & \frac{1}{16} & \frac{1}{2} & \frac{1}{40} & \frac{1}{5} & \frac{1}{64} & 0 & \frac{1}{64} & 0 & \frac{1}{64} & 0 & \frac{1}{64} \\ 
\e\in B & n/a & 0 & 1 & \frac{1}{28} & \frac{2}{3} & \frac{3}{112} & \frac{2}{5} & \frac{1}{56} & \frac{4}{35} & \frac{1}{64} & 0 & \frac{1}{64} & 0 & \frac{1}{64} & 0 & \frac{1}{64}  \\ 
\e\in C & n/a & 0 & n/a & 0 & \frac{1}{3} & \frac{1}{112} & \frac{1}{4} & \frac{1}{70} & \frac{2}{35} & \frac{1}{64} & 0 & \frac{1}{64} & 0 & \frac{1}{64} & 0 & \frac{1}{64}  \\ 
\end{array}
\end{equation}

One can conclude from this data that if one measures a syndrome of, say $(110011)\in C$, then the error correction will most likely fail.

We also computed the probabilities that Shor's code will correct an error affecting $t$ qubits after the syndrome $\e$ is measured.  We omit the results, but mention that these probabilities are constant for error syndromes lying in each of six different sets.

\section{Conclusion}

Even in a classical $[n,k,d]$ code, there are circumstances when $t$ bitflips can be corrected when $2t\geq d.$  However, for quantum codes the situation is made more interesting by the probabilistic quantum measurement.   Finer analysis beyond noting the invariants $n$, $k$, and $d$, may be necessary to accurately judge the effectiveness of the code.  
As illustrated here, 
whether the nine qubit code or the smaller seven qubit code is better depends on the value of $p$ in the error model.

During the actual implementation of the code, one should consider the syndrome quality.  In advance, a tolerable threshold for syndrome quality can be chosen.  Then, once a syndrome is measured, a lookup can be performed to determine if the error correction procedure will succeed with a probability less than the tolerable threshold.  If not, the error correction is aborted.  This kind of distillation may have many applications.  It is feasible, for instance, that the accuracy threshold for fault tolerant computing could be lowered by considering probabilistic error correction, and could be lowered further by accounting for syndrome quality, at a negligible resource cost.

\end{document}